\documentclass[sn-mathphys-ay, Namedate, iicol]{sn-jnl}
\usepackage{graphicx}    
\usepackage{amsmath}     
\usepackage{amssymb}     
\usepackage{amsfonts}    
\usepackage{xspace}
\usepackage{amsthm}%
\usepackage{mathrsfs}%
\usepackage[title]{appendix}%
\usepackage{xcolor}%
\usepackage{textcomp}%
\usepackage{manyfoot}%
\usepackage{booktabs}%
\usepackage{listings}%
\newcommand{\Msun}{\,$M_{\odot}$\xspace}
\newcommand{\Rsun}{\,$R_{\odot}$\xspace}
\newcommand{\Msyr}{\,$M_{\odot}$\,yr$^{-1}$\xspace}
\newcommand{\kms}{\,km\,s$^{-1}$\xspace}
\newcommand{\ergs}{\,erg\,s$^{-1}$\xspace}
\newcommand{\gcm}{\,g\,cm$^{-1}$\xspace}
\newcommand{\gcmq}{\,g\,cm$^{-3}$\xspace}

\newcommand{\Ha}{H$\alpha$\xspace}
\newcommand{\Hb}{H$\beta$\xspace}
\newcommand{\Hg}{H$\gamma$\xspace}

\newcommand{\HeI}{He\,I\xspace}
\newcommand{\HeII}{He\,II\xspace}

\newcommand{\FeII}{Fe\,II\xspace}
\newcommand{\A}{\,\AA\xspace}
\newcommand{\apj}{Astrophys. J. } 
\newcommand{\apjl}{Astrophys. J. Lett. } 
\newcommand{\apss}{Astrophys. Space Sci. } 
\newcommand{\aap}{Astron. Astrophys. } 
\newcommand{\fcp}{Fundamentals of Cosmic Physics } 
\newcommand{\mnras}{Mon. Not. R. Astron. Soc. } 
\begin{document}
\title[Short-plateau SN~2018gj] 
{Revisiting short-plateau SN~2018gj}

\author*[1,2]{\fnm{V. P.} \sur{Utrobin}}\email{utrobin@itep.ru}

\author[2]{\fnm{N. N.} \sur{Chugai}}\email{nchugai@inasan.ru}


\affil*[1]{\orgname{NRC ``Kurchatov Institute''},
   \orgaddress{\street{acad. Kurchatov Square 1}, \city{Moscow},
   \postcode{123182}, \country{Russia}}}

\affil[2]{\orgname{Institute of Astronomy, Russian Academy of Sciences},
   \orgaddress{\street{Pyatnitskaya St. 48}, \city{Moscow},
   \postcode{119017}, \country{Russia}}}

\abstract{%
We present an alternative model of unusual type-IIP SN~2018gj.
Despite the short plateau and early gamma-rays escape seeming to favor low-mass
   ejecta, our hydrodynamic model requires a large ejected mass
   ($\approx$23\Msun).
The high ejecta velocity, we find from hydrogen lines in early spectra,
   is among crucial constraints on the hydrodynamic model.
We recover the wind density that rules out a notable contribution of
   the circumstellar interaction to the bolometric luminosity.
The early radioactive gamma-rays escape is found to be due to
   the high velocity of $^{56}$Ni, whereas the asymmetry of the \Ha emission
   is attributed to the asymmetry of the $^{56}$Ni ejecta.
The available sample of type-IIP supernovae studied hydrodynamically in
   a uniform way indicates that the asymmetry of the $^{56}$Ni ejecta is
   probably their intrinsic property.
Hydrogen lines in the early spectra of SN~2018gi and SN~2020jfo are found to imply
   a clumpy structure of the outer ejecta.
With two already known similar cases of SN~2008in and SN~2012A we speculate
   that the clumpiness of the outer ejecta is inherent to type-IIP supernovae
   related to the red supergiant explosion.
}


\maketitle

\section{Introduction} 
\label{sec:intro}
Type-IIP and -IIL supernovae (SNe IIP/L) compose the dominant category of
   core-collapse supernovae that originate from progenitors with
   the main-sequence masses of $9-25$\Msun \citep{Woosley_2002}. 
The unbiased sample of 13 SNe~IIP explored via the uniform hydrodynamic approach
   \citep{UC_2024} is roughly consistent with the predicted mass range of
   SNe~IIP/L progenitors, although ejecta masses $\lesssim$12\Msun
   are somewhat scarce in this sample.

However, the modeling of SN~2020jfo with a short light-curve plateau ($\sim$60\,d)
   results in the ejecta mass of $\approx$6\Msun \citep{Teja_2022, UC_2024}.
Moreover, \cite{Hiramatsu_2021} reported on three SNe~IIP with short plateaus
   (SN~2008Y, SN~2006ai, and SN~2016egz) and the inferred hydrogen envelope
   masses of about 1, 2, and 4\Msun, respectively.
Hence, the low-mass ejecta --- although rare among SNe~IIP --- are not extremely scarce.

The type-IIP supernova SN~2018gj in the galaxy NGC 6217 discovered by
   \cite{Wiggins_2018} and explored in detail by \cite{Teja_2023} is the
   short-plateau ($\sim$75\,d) object.
Another notable feature is a rapid luminosity decline at the radioactive tail
   just after the plateau stage.
At first glance, both the short plateau and the early gamma-ray leakage suggest
   a low-mass ejecta that seems to be in line with the hydrodynamic modeling
   \citep{Teja_2023}.
However, obvious dissimilarities between SN~2018gj and SN~2020jfo ---  longer
   plateau of the former (75 vs 60\,d), higher plateau luminosity,  \Ha\ emission
   asymmetry, and early gamma-ray leakage --- suggest that SN~2018gj is
   a special case and not just a twin of SN~2020jfo.
The above said motivates us to revisit the case of SN~2018gj.

There are two additional important reasons for this.
First, the previous hydrodynamic model of SN~2018gj requires $0.1-0.2$\Msun
   in the circumstellar (CS) shell to account for the early luminosity peak
   \citep{Teja_2023}.
Meanwhile, the issue of the CS matter density can be explored based on
   available spectra in a way similar to that of SN~2020jfo \citep{UC_2024},
   though with some correction.
For SN~2018gj, the early spectra at $1-3$ days are absent, so we cannot use
   the early-time broad \HeII\,4686\,\AA\ emission to infer the maximal
   expansion velocity.
However, we can use hydrogen lines at a somewhat later stage and obtain
   the maximal velocity from the absorption blue edge.

Secondly, the early spectrum of SN~2018gj provides us with the opportunity
   to verify a possible ubiquity of the clumpy structure of the outer ejecta
   of SNe~IIP.
This phenomenon is revealed by SN~2008in \citep{CU_2014} and SN~2012A
   \citep{UC_2015} via the anomalously low ratio of optical depths \Ha/\Hb
   ($\sim$1, instead of a theoretical ratio of 7.25); we dub this paradox
   the ``\Ha/\Hb-problem''.  
The \Ha/\Hb-problem is resolved assuming the clumpy outer ejecta \citep{CU_2014}.
The ejecta clumpiness was attributed to the propagation of the explosion
   shock wave in the red supergiant (RSG) envelope, perturbed before that
   by a vigorous convection.
The issue of a shock breakout (SBO) in the RSG envelope modified by the convection
   has recently received attention and has been explored via three-dimensional (3D)
   simulations \citep{Goldberg_2022b, Goldberg_2022a}.

In Section~\ref{sec:hvel}, we recover the maximal expansion velocities via
   the modeling of hydrogen lines that also reveals the \Ha/\Hb-problem.
We then find the basic parameters of SN~2018gj via the hydrodynamic modeling
   of the relevant set of observational data including the maximal ejecta
   velocity (Section~\ref{sec:model}).
In Section~\ref{sec:wind}, we infer the density of the presupernova (pre-SN)
   wind and show that the ejecta interaction with the CS matter cannot affect
   the light curve.
Finally, we discuss the results and some features of the SNe~IIP sample studied
   with a uniform hydrodynamic approach.

Below, we use the distance modulus $\mu = 31.46$\,$\pm$\,$0.15$ ($D = 19.61$\,Mpc),
   the reddening $E(B-V) = 0.08$\,$\pm$\,$0.02$\,mag,
   and the explosion date of JD 2458127.8 \citep{Teja_2023}.

\section{High-velocity ejecta}
\label{sec:hvel}
%
\subsection{Outermost ejecta velocity}
\label{sec:hvel-outermost}
The parameters of SN~IIP --- ejecta mass, explosion energy, and pre-SN radius ---
   can be reliably inferred from the light curve and expansion velocities
   via modeling the well-observed SNe based on all sensitive observables.
Particular attention should be paid to the maximum ejecta velocity that is
   associated with the thin boundary dense shell formed during the SBO
   \citep{GIN_1971, Chevalier_1981}.
Usually, the hydrodynamic modeling of SNe~IIP ignores the ejecta velocities at
   the early stage ($t < 10$\,d) focusing primarily on the photospheric
   velocities at the plateau stage.
However, it is well known that SN parameters inferred from the description of
   the SN~IIP plateau luminosity and the expansion velocities at the plateau
   stage suffered from parameter degeneracy \citep{Goldberg_2020}. 
On the other hand, taking into account the outermost ejecta velocity secures
   the unique choice of principal SN parameters \citep{UC_2019}.

Another key role of the maximal ejecta velocity is the constraining of
   the CS density, which permits us to rule out a dense CS shell frequently
   invoked to describe the early luminosity peak.
Indeed, a high ejecta velocity indicates the absence of significant deceleration
   and thus excludes a dense CS shell.
The case of SN~2020jfo illustrates the efficiency of this observational constraint
   \citep{UC_2024}.

The available set of SN~2018gj spectra \citep{Teja_2023} starts from day 5
   after the SN explosion.
The hydrogen and helium lines with a shallow broad P Cygni line profile are seen
   already on day 5, while the \Ha line becomes pronounced on day 7.
We recover the velocity at the  photosphere ($v_p$) and the maximum velocity of
   the ejecta ($v_{max}$) for hydrogen and \HeII\,5876\A lines using the spectrum
   on day 7.

The line profile is modeled assuming a spherical atmosphere attached to
   a photosphere with the sharp boundary.
The Sobolev optical depth $\tau$ and emissivity $\eta$ are set parametrically as
   $\tau = \tau_p(v_p/v)^3$ and $\eta = \eta_p(v_p/v)^q$, where $q$ is in the range
   of $6-9$.
The radiation transfer is calculated using the Monte Carlo technique.
The photon with the weight $w$ scatters at the resonant point $v$ with the
   probability of $(1 - e^{-\tau})$; the scattered photon acquires the
   weight $\lambda w$ with the albedo $\lambda = 1$ for \Ha, while $\lambda < 1$
   for \Hb and \Hg due to the transition to lower levels.
The net line emission is described as the term $\eta w$ added to the weight of
   scattered photons.
The parameter of net emissivity is in the range of $0 < \eta_p <4$ with
   the maximum value for \Ha.

The line photon striking the photosphere can be reflected diffusively, which
   results in a blueshift compared to the case of absorbing photosphere
   (Fig.~\ref{fig:clumpy}b, inset).
With the density at the photosphere provided by the hydrodynamic model
   $\approx$10$^{-13}$\gcmq on day 7 and the observational temperature of
   $\approx$10\,000\,K \citep{Teja_2023}, the free-free absorption is the major
   absorption mechanism at the photosphere.
Using Boltzmann-Saha equations for ionization and excitation, we estimate
   the absorption probability ($\epsilon$) in the range of $0.002-0.007$
   for wavelengths between \Hg and \Ha.
The spherical albedo in the case of the high optical depth and $\epsilon < 0.2$
   can be approximated as $A \approx 1 -2\sqrt{\epsilon}$ \citep{Sobolev_1975}.
This results in $A \approx 0.83-0.91$ which means that the diffusive reflection
   must be taken into account.

\begin{table}
\centering 
\caption{The photospheric and maximal velocities and the Sobolev optical depth
   in SN~2018gj and SN 2020jfo ejecta on day 7}
\label{tab:vel}
\begin{tabular}{l c c c c}
\toprule
\noalign{\smallskip}
SN      &  Line &  $v_p$  &  $v_{max}$ & $\tau_p$ \\
\noalign{\smallskip}
\midrule
\noalign{\smallskip}       
2018gj  &  \Hg  &  $9.5\pm0.5$ & $15.2\pm1$  & $0.23\pm0.05$ \\
        &  \Hb  &  $11\pm0.5$  & $16.5\pm1$  & $0.27\pm0.05$ \\
        &  \HeI &  $10\pm0.5$  & $15.5\pm1$  & $0.25\pm0.05$ \\
        &  \Ha  &  $10\pm0.5$  & $15\pm1$    & $0.27\pm0.05$ \\        
\noalign{\smallskip}       
\hline
\noalign{\smallskip}       
2020jfo &  \Hg  &  $9.5\pm0.5$ & $15.2\pm1$  & $0.2\pm0.05$  \\
        &  \Hb  &  $9.5\pm0.5$ & $16.15\pm1$ & $0.4\pm0.05$  \\
        &  \HeI &  $9\pm0.5$   & $15.3\pm1$  & $0.25\pm0.05$ \\
        &  \Ha  &  $9\pm0.5$   & $15.3\pm1$  & $0.33\pm0.05$ \\   
\botrule
\end{tabular}
\footnotetext{Velocities are measured in 1000\kms.}
\end{table}
The recovered velocities and the Sobolev optical depth at the photosphere
   on day 7 are given in Table~\ref{tab:vel}.
The parameters of SN~2020jfo with the shorter plateau are included for comparison.
The uncertainties shown in Table~\ref{tab:vel} are typical scatter of
   the results that is estimated via parameter variations with a visual model
   fitting in each case and then rounded.
The mean maximal velocity for the considered lines of SN~2018gj is
   $v_{max} = 15\,500$\,$\pm$\,$500$\kms.
Surprisingly, the inferred values for both SNe are similar.
In the case of SN~2020jfo, we obtained the maximal velocity of 16\,500\kms
   from the broad \HeII\,4686\A emission on day 2.1 \citep{UC_2024}.
Although the early spectra are lacking for SN~2018gj, the inferred similar
   maximal velocity on day 7 in both SNe suggests that the maximal velocity
   on day 2.1 for SN~2018gj could be comparable to that of SN~2020jfo,
   assuming that the rate of the velocity evolution is the same for both
   SNe.

\subsection{Clumpiness of outer ejecta}
\label{sec:hvel-clumpiness}
%
\begin{figure}
   \includegraphics[width=\columnwidth, clip, trim=46  86  41 110] {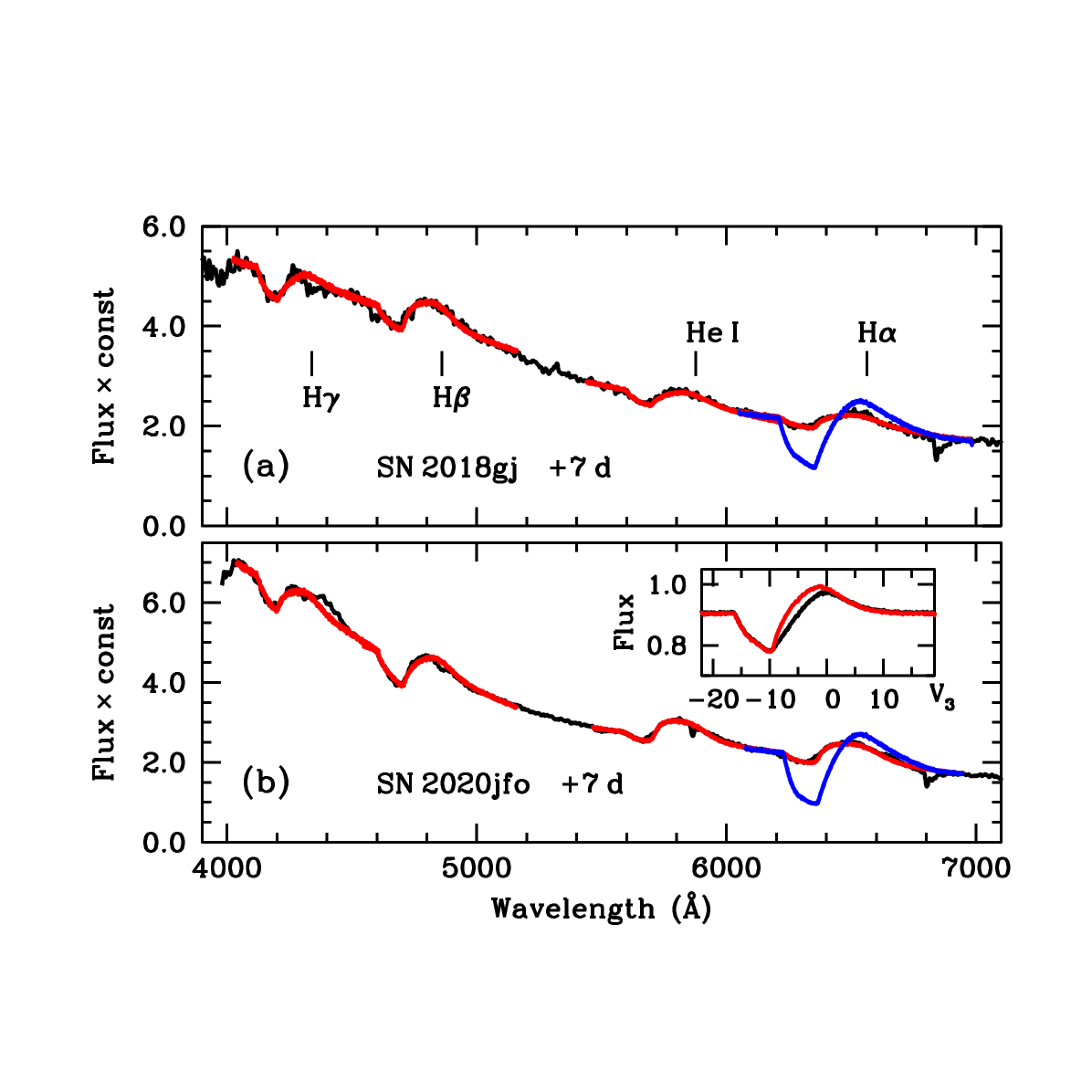}
   \caption{%
   Spectra of SN~2018gj (Panel (a)) and SN~2020jfo (Panel (b)) on day 7
      showing \Hg, \Hb, \HeI\,5876\AA, and \Ha lines with the overlaid line
      models (\emph{red line\/}).
   For the \Ha line we show a case with the theoretical ratio of the \Ha/\Hb
      optical depths (\emph{blue line\/}); this illustrates the ``\Ha/\Hb
      problem'' related to the clumpiness (see Section~\ref{sec:hvel-clumpiness}). 
   Inset in the lower panel displays the \Hb line profile for the photosphere
      albedo $A = 1$ (\emph{red line\/}) and $A = 0$ (\emph{black line\/})
      that demonstrates the effect of the diffuse reflection from
      the photosphere.
   }
   \label{fig:clumpy}
\end{figure}
A striking feature is a comparable value of the Sobolev optical  depth for
   \Ha and \Hb demonstrated by both SNe.
This fact is in sharp contradiction with the theoretical ratio 
   $\tau_{23}/\tau_{24} = 7.25$. 
The \Ha/\Hb problem is clearly demonstrated in Fig.~\ref{fig:clumpy} that
   shows the unacceptably deep \Ha absorption expected theoretically
   based on the \Hb optical depth.

In fact, this phenomenon has been already recognized in type-IIP SN~2008in
   \citep{UC_2013} and explored in terms of the ejecta clumpy structure
   \citep{CU_2014}.
The effect of clumpiness has been found also in SN~2012A \citep{UC_2015}.
It is reasonable to assume that in SN~2018gj and SN~2020jfo the outer
   ejecta are composed by dense clumps that are optically thick ($\tau \gg 1$)
   in hydrogen lines, whereas the rest of the volume is optically thin
   ($\tau \ll 1$) for the same lines.
In that case, the absorbed (scattered) fraction of the photosphere flux
   at a certain radial velocity is determined by the occultation optical depth, 
   or average volume filling factor ($f$),
   at the resonant plane\footnote{For the space filled by bodies
   with the filling factor $f$ the relative area of body sections
   in the random plane is also $f$.} and not by the Sobolev optical depth
   as expected for the homogeneous density distribution.
Remarkably, the clumpy structure of the outer ejecta in SN~2008in, SN~2012A,
   SN~2018gj, and SN~2020jfo is characterized by the similar filling factor
   ($f \sim 0.3$), which apparently is the intrinsic property of a mechanism
   responsible for the clumpiness.

The clumpy structure presumably arises during the explosion shock-wave propagation in
   the outer layers of the RSG star where a vigorous convective
   motion with the convective velocity comparable to the sonic speed
   $v_c \sim c_s$ bring about high-amplitude density perturbations
   $\Delta \rho/\rho \approx (v_c/c_s)^2 \sim 1$ \citep{CU_2014}.
These density fluctuations affect the shock-wave propagation, thus producing
   the clumpiness of the outer ejecta.
It goes without saying that the one-dimensional (1D) hydrodynamics of SNe~IIP is not
   able to reproduce the clumpiness of the outer ejecta, however, 3D hydrodynamics
   hopefully could do so in the future \citep[cf.][]{Goldberg_2022b}.

\section{Hydrodynamic model}
\label{sec:model}
%
\subsection{Model overview}
\label{sec:model-overview}
We use the radiation hydrodynamics code {\sc CRAB} with the radiation transfer
   in the gray approximation \citep{Utrobin_2004, Utrobin_2007}.
The pre-SN is the hydrostatic nonevolutionary RSG model.
The term ``nonevolutionary'' means that the pre-SN density distribution and
   the chemical composition are modified in order to reproduce the light curve
   and expansion velocities.
There are two reasons to use the nonevolutionary model: (i) the explosion of
   an evolutionary model in a 3D hydrodynamics results in the strong modification
   of the pre-SN density and composition distributions in the hydrodynamic time
   scale; (ii) the explosion of the evolutionary RSG star using the
   3D hydrodynamics is not able to reproduce the light curve of the standard
   type-IIP SN~1999em \citep{Utrobin_2017}.
These arguments compel us to use a nonevolutionary model that should be
   considered as a palliative tool to compensate for the lack of an adequate
   ``ab initio'' model with the appropriate physics included.

\begin{figure}
   \includegraphics[width=\columnwidth, clip, trim=0 237 53 139]{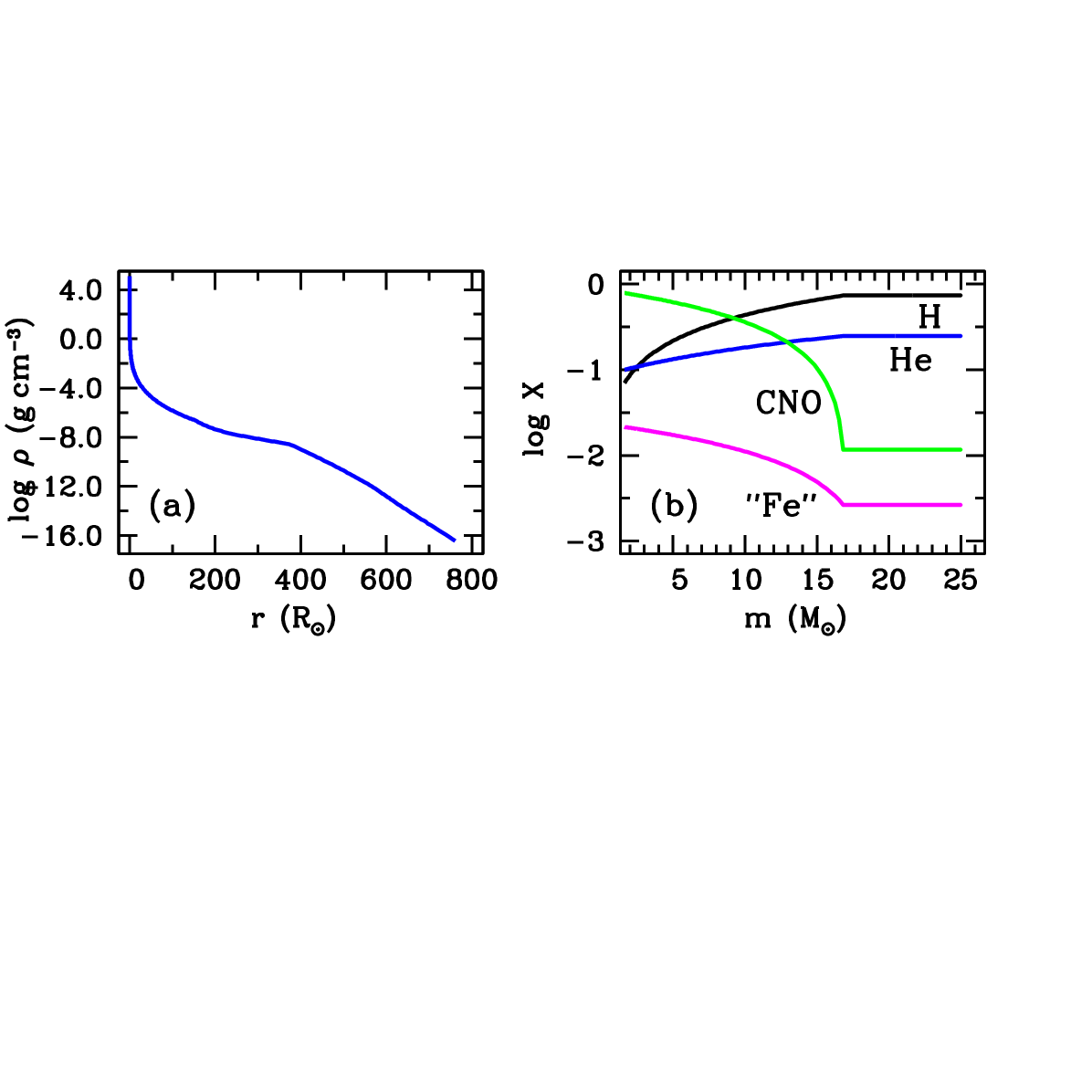}
   \caption{%
   The structure of the pre-SN model.
   Panel (a): the density distribution as a function of radius.
   Panel (b): the chemical composition.
   Mass fraction of hydrogen (\emph{black line\/}), helium
      (\emph{blue line\/}), CNO elements (\emph{green line\/}),
      and Fe-peak elements excluding radioactive $^{56}$Ni
      (\emph{magenta line\/}) in the ejected envelope.
   The central core of 1.6\Msun is omitted.
   }
   \label{fig:presn}
\end{figure}

The explosion is initiated by a supersonic piston applied to the stellar
   envelope at the boundary with the 1.6\Msun collapsing core.
The description of the light curve and velocities at the photosphere,
   including the outermost expansion velocity, requires a tuning of
   the pre-SN density and chemical-composition distributions
   that should have smooth density and composition gradients
   at the metals/He and He/H interfaces (Fig.~\ref{fig:presn}).
The smoothed gradients presumably reflect mixing in 3D simulations of
   the SN~IIP explosion \citep{Utrobin_2017}.

\begin{table}
\centering 
\caption{Parameters of hydrodynamic model}
\label{tab:param}
\begin{tabular}{l c c c}
\toprule
\noalign{\smallskip}
Parameter & Unit & Value & Error \\
\noalign{\smallskip}
\midrule
\noalign{\smallskip}
Ejected mass     & \Msun          & 23.4  & $\pm$\,1.6   \\
Explosion energy & $10^{51}$\,erg & 1.84  & $\pm$\,0.14  \\
Pre-SN radius    & \Rsun	      & 775   & $\pm$\,55    \\
$^{56}$Ni mass   & \Msun          & 0.031 & $\pm$\,0.005 \\
\botrule
\end{tabular}
\end{table}
%
\subsection{Results}
\label{sec:model-results}
\begin{figure}
   \includegraphics[width=\columnwidth, clip, trim=0 239  53 134]{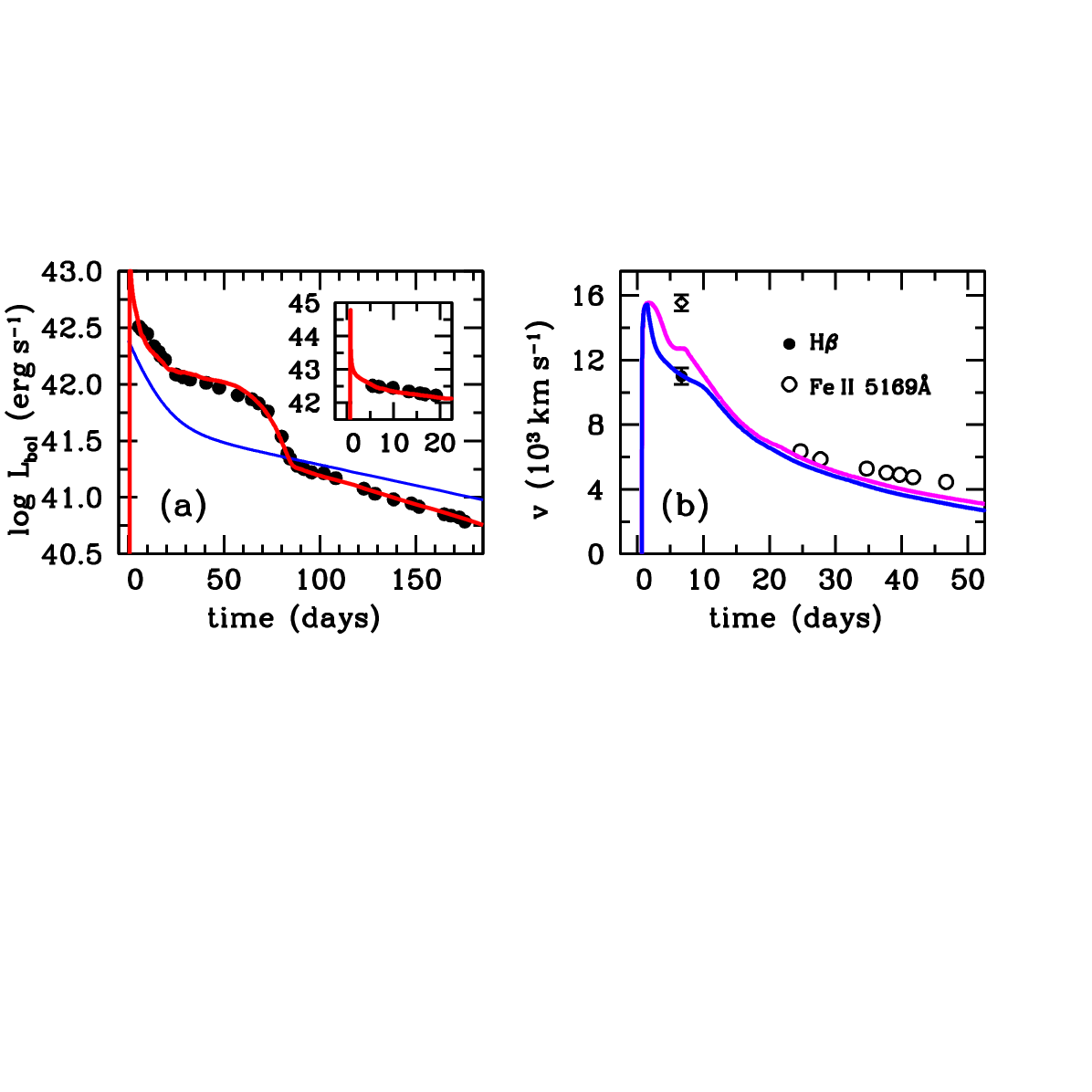}
   \caption{%
   The bolometric light curve and the evolution of photospheric velocity.
   Panel (a): the model light curve (\emph{red line\/}) overlaid on
      the bolometric data (\emph{circles\/}) \citep{Teja_2023}.
   The \emph{blue line} is the total power of radioactive $^{56}$Ni decay.
   Inset shows zoom-in of the initial 20 days.
   Panel (b): the evolution of model velocity defined by the level
      $\tau_{eff} = 2/3$ (\emph{blue line\/}) and $\tau_\mathrm{Thomson} = 1$
      (\emph{magenta line\/}) is compared with the photospheric velocities
      estimated from the absorption minimum of \FeII 5169\A \citep{Teja_2023}
      along with our estimate from the \Hb line.
   The outermost ejecta velocity is recovered from the blue absorption wing of
      hydrogen lines (\emph{diamond\/}).
   }
   \label{fig:lcv}
\end{figure}

The hydrodynamic model with parameters listed in Table~\ref{tab:param} provides
   an optimal fit to the bolometric light curve together with the velocity at
   the photosphere (Fig.~\ref{fig:lcv}).
The model maximal velocity of 15\,800\kms is also in reasonable agreement with
   $v_{max} = 15\,550$\,$\pm$\,500\kms inferred above from the hydrogen absorption line
   profiles in the spectrum on day 7.
The missing rising  part of the light curve hampers the recovery of the explosion
   moment with accuracy better than 1.5\,d.
The overall fit of the light curve and the expansion velocities
   (Fig.~\ref{fig:lcv}) agree with the adopted explosion date JD 2458127.8
   \citep{Teja_2023}.

\begin{figure}
   \includegraphics[width=\columnwidth, clip, trim=0 239 53 139]{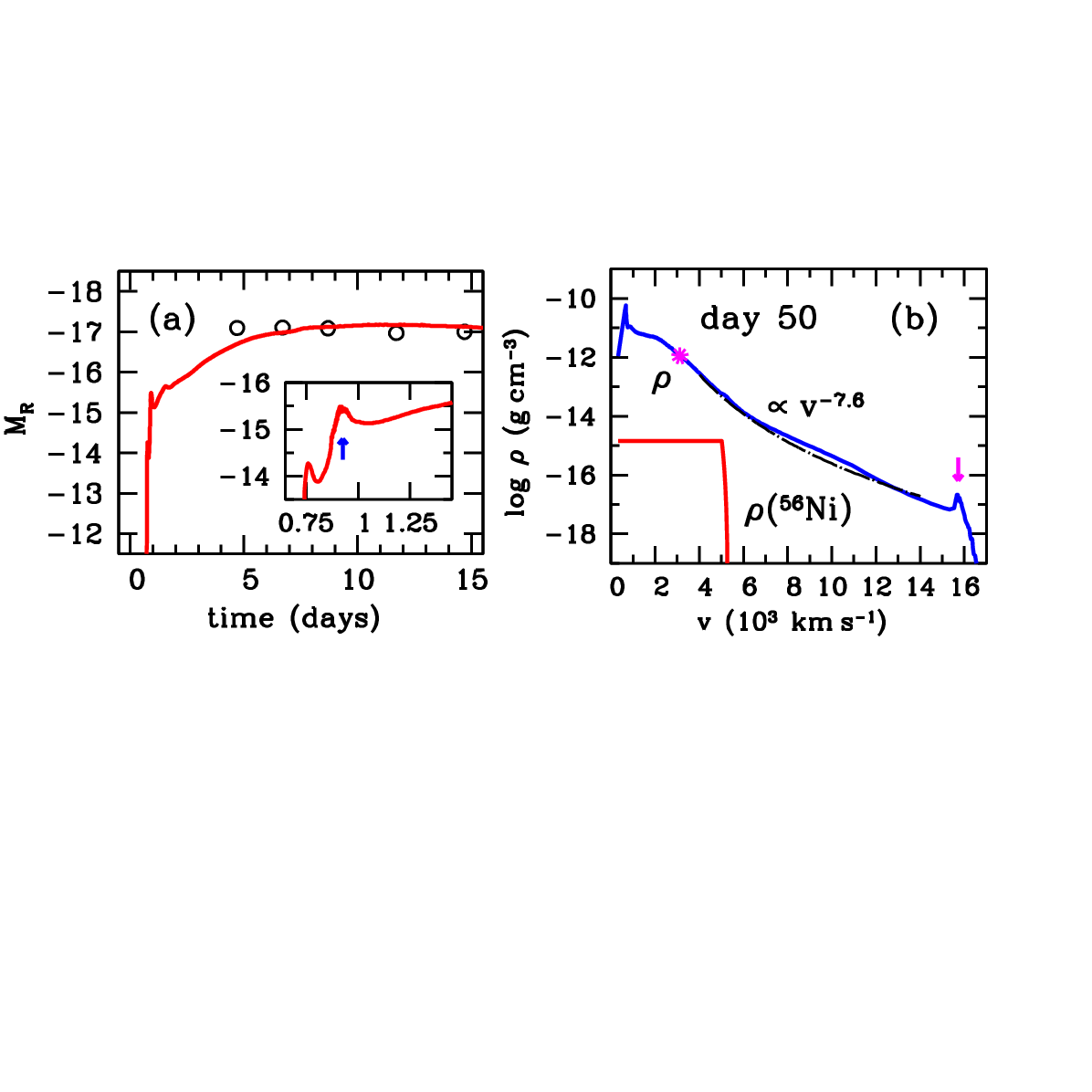}
   \caption{%
   Panel (a): Rising part of the model light curve in the $r$-band overplotted
      on the observational data taken by \citet{Teja_2023}.
   Inset shows the fine structure of the narrow peak related to shock breakout.
   The second fine-structure peak indicated by blue arrow corresponds to
      the formation of the thin boundary shell marked by an arrow
      on the right panel.
   Panel (b): The density and $^{56}$Ni distributions vs. velocity in the ejecta
      on day 50; magenta star indicates the photosphere location, while the magenta
      arrow shows the boundary thin shell.
   The \emph{dash-dotted} line is the power law $\rho \propto v^{-7.6}$.
   }
   \label{fig:rise}
\end{figure}

Despite the fact that the radiation transfer is treated in the gray approximation, the early
   $R$-band magnitude is well reproduced by the model (Fig.~\ref{fig:rise}a).
As in the previous SNe~IIP models, the initial $R$-band peak has
   the double structure that was explained in the case of SN~2017gmr
   \citep{UC_2021}.
The first peak is related to the SBO, whereas the second is the outcome of
   the thin-shell formation.
However, it is not clear whether this double-peak structure would remain in
   3D hydrodynamics, since density perturbations in the RSG envelope
   modify the luminosity peak following the SBO \citep{Goldberg_2022b}.

The total density and the $^{56}$Ni density in the freely expanding ejecta
   on day 50 are shown in Fig.~\ref{fig:rise}b.
The remarkable result of the SN~2018gj modeling is the high velocity of
   the $^{56}$Ni ejecta ($\approx$5280\kms) that is imposed by the rapid
   luminosity decline at the radioactive tail (Fig.~\ref{fig:lcv}a).
This behavior of the light curve also suggests the total $^{56}$Ni mass of
   0.031\Msun, somewhat larger than the previous estimate of 0.026\Msun
   \citep{Teja_2023}.
The inferred $^{56}$Ni velocity is significantly larger compared to another
   short-plateau SN~2020jfo, where the $^{56}$Ni ejecta is located in
   the center with the maximum velocity of 1600\kms \citep{UC_2024}.
The \Ha emission blueshift \citep{Teja_2023} is likely
   an effect of the bipolar $^{56}$Ni ejecta, whereas the 1D hydrodynamical model
   treats the $^{56}$Ni ejecta as spherically symmetrical.
The inferred $^{56}$Ni velocity therefore should be considered as the maximal
   velocity of the asymmetric $^{56}$Ni distribution.

The uncertainty in the derived SN parameters can be estimated by a variation
   of the model parameters around the optimal model.
The uncertainties of the distance and the reddening (see Section~\ref{sec:intro})
   imply the 15 per cent uncertainty in the bolometric luminosity.
The scatter in the plot of the photospheric velocity versus time
   (Fig.~\ref{fig:lcv}b) suggests an uncertainty of 6 per cent
   in the photospheric velocity.
Following \citet{Teja_2023}, we adopt the uncertainty of the plateau length
   as 2\,d, i.e., 2.7 per cent of the plateau duration.
With these uncertainties of observables, we find errors of
   $\pm$\,55\Rsun for the initial radius, $\pm$\,1.6\Msun for the ejecta
   mass, $\pm$\,$0.14\times10^{51}$\,erg for the explosion energy, and
   $\pm$\,0.005\Msun for the total $^{56}$Ni mass (Table~\ref{tab:param}).

\section{Presupernova wind} 
\label{sec:wind}
%
\begin{figure}
   \includegraphics[width=\columnwidth, clip, trim=18 114  28 116]{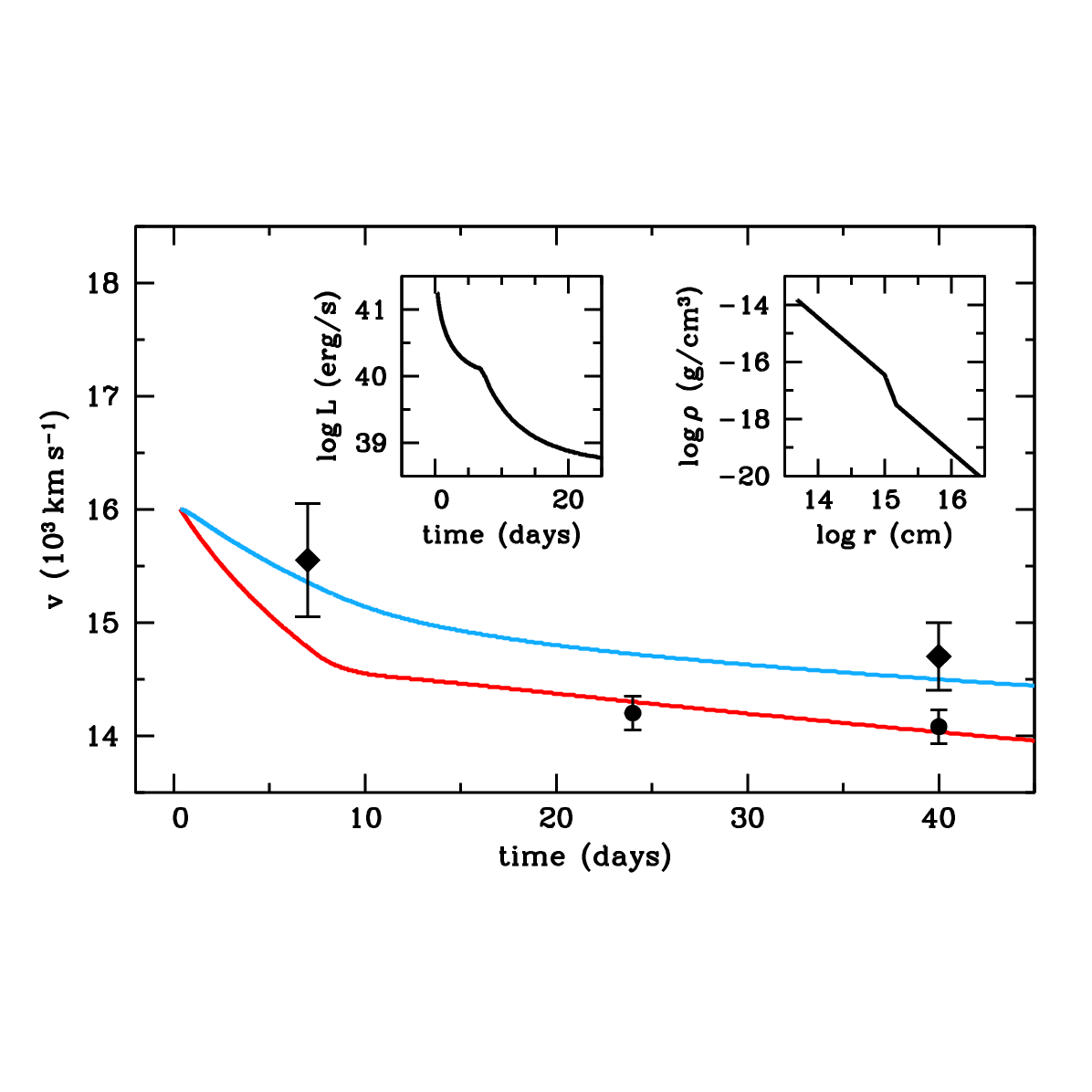}
   \caption{%
   The CDS velocity (\emph{red line\/}) and the boundary velocity of the unshocked
      ejecta (\emph{blue line\/}) in the model of the SN/wind interaction
      overlaid on the observed velocities of the CDS (\emph{circles\/}) and
      the unshocked ejecta (\emph{diamonds\/}).
   The left inset shows the combined luminosity of the forward and reverse shocks;
      the right inset shows the wind density.
   }
   \label{fig:wind}
\end{figure}
The density and radial distribution of the CS matter can be recovered from
   the CS interaction effects imprinted in spectra of SN~2018gj \citep{Teja_2023}.
The hydrodynamics of the ejecta interaction with the CS gas is considered
   in the thin-shell approximation \citep{Chevalier_1982, Chugai_2018}.
The major observables that constrain the CS density distribution are the maximal
   velocity of the unshocked ejecta inferred from four lines on day 7,
   the \Ha line on day 40, and also the high-velocity narrow absorption (HVNA)
   in the blue \Ha wing on days 24 and 40 that indicate the velocity of
   the cold dense shell (CDS) \citep{Chugai_2007}.
The density distribution in the ejecta is approximated by the expression
   $\rho = \rho_0(t_0/t)^3/[1 + (v/v_0)^{7.6}]$, where $\rho_0$ and $v_0$ are
   defined by the ejecta parameters ($M$ and $E$).

We find that the steady wind $\rho \propto r^{-2}$ with density parameter
   $w = \dot{M}/u =2\times10^{14}$\gcm matches the boundary velocity of
   the unshocked ejecta, but does not fit the CDS velocity.
The appropriate model (Fig.~\ref{fig:wind}) suggests an enhanced wind density
   at radii $r < 1.5\times10^{15}$\,cm and a steady wind at
   $r > 1.5\times10^{15}$\,cm with the density parameter
   $w = \dot{M}/u =8.7\times10^{13}$\gcm.
This wind corresponds to the mass-loss rate $\dot{M} = 2\times10^{-6}u_{15}$\Msyr
   assuming the wind speed $u_{15} = u/15$\kms$=1$.
 
The combined luminosity of the forward and reverse shocks at the maximum
   is $\approx$2$\times10^{41}$\ergs, lower by two orders than the maximal
   bolometric luminosity; a similar difference remains at later epochs.
The luminosity related to the CS interaction in SN~2018gj thus is negligibly small.
This conclusion is in line with the low mass of the CS shell
   ($3\times10^{-4}$\Msun) within the radius $r < 1.5\times10^{15}$\,cm,
   lower by almost three orders than the amount required to maintain
   the initial luminosity peak of SN~2018gj due to the CS interaction.

\section{Discussion}
\label{sec:disc}
%
\begin{table}
\centering
\caption{Hydrodynamic models of type-IIP supernovae}
\label{tab:sumtab}
\begin{tabular}{@{ } l  @{ } c  @{ } c @{ } c @{ } c @{ } c @{ } c}
\toprule
\noalign{\smallskip}
 SN & $R_0$ & $M_{ej}$ & $E$ & $M_{\mathrm{Ni}}$ & $v_{\mathrm{Ni}}^{max}$ & S/A \\
    & (\Rsun) & (\Msun) & ($10^{51}$\,erg) & (\Msun) & (km\,s$^{-1}$) & \\
\noalign{\smallskip}
\midrule
\noalign{\smallskip}
 1987A  &  35  & 18   & 1.5    & 0.0765 &  3000 & A \\
1999em  & 500  & 19   & 1.3    & 0.036  &  660  & A \\
2000cb  &  35  & 22.3 & 4.4    & 0.083  &  8400 & A \\
 2003Z  & 230  & 14   & 0.245  & 0.0063 &  535  & S \\
2004et  & 1500 & 22.9 & 2.3    & 0.068  &  1000 & S \\
2005cs  & 600  & 15.9 & 0.41   & 0.0082 &  610  & S \\
2008in  & 570  & 13.6 & 0.505  & 0.015  &  770  & S \\
2009kf  & 2000 & 28.1 & 21.5   & 0.40   &  7700 & -- \\
2012A   &  715 & 13.1 & 0.525  & 0.0116 &  710  & S \\
2013ej  & 1500 & 26.1 & 1.4    & 0.039  &  6500 & A \\
 2016X  &  436 & 28.0 & 1.73   & 0.0295 &  4000 & A \\
2017gmr &  525 & 22.0 & 10.2   & 0.110  &  3300 & A \\
2018gj  &  775 & 23.4 & 1.84   & 0.031  &  5280 & A \\
2020jfo &  400 &  6.2 & 0.756  & 0.013  &  1600 & S \\
\botrule
\end{tabular}
\footnotetext{The last column indicates a degree of asymmetry in the $^{56}$Ni
   ejecta. the ``S'' symbol stands for seemingly symmetric distribution and
   the ``A'' symbol denotes an apparent asymmetry.
   SN~2009kf has no clear signature for the $^{56}$Ni geometry.}
\end{table}
The massive ejecta --- the major outcome of the hydrodynamic modeling of
   SN~2018gj --- prima facie seems odd, given the short plateau that presumably
   should be associated with a low-mass ejecta. 
Particularly impressive is the comparison with SN~2020jfo, another
   short-plateau SN~IIP with the ejecta mass of $\sim$6\Msun
   \citep{Teja_2022, UC_2024}.
The point, however, is that the plateau duration depends not only on the ejecta
   mass, but also on the explosion energy, the pre-SN radius, the amount of
   $^{56}$Ni, and its velocity.
The inferred SN~2018gj parameters are fixed by the whole set of the relevant
   observables (the bolometric light curve and expansion velocities).

It should be emphasized that the crucial observables include --- apart from
   the light curve and the photospheric velocities --- the maximal velocity of
   the ejecta usually ignored in the hydrodynamic description of SN~IIP.
It is noteworthy that we recover the velocities of the external ejecta from the
   blue absorption wing of hydrogen lines, so these velocities somewhat exceed
   the velocity at the photosphere.
  
An interesting byproduct of the spectral analysis is the demonstration that
   the structure of the external high-velocity ejecta of SN~2018gj and SN~2020jfo
   is essentially clumpy.
The clumpiness of the outer ejecta seems to be a common feature of the ordinary
   SNe~IIP, since the clumpy structure of the external ejecta has been also
   recovered in type-IIP SN~2008in \citep{CU_2014} and SN~2012A \citep{UC_2015}.
Such a structure of the outer ejecta of SNe~IIP suggests that the shock
   propagation during the SBO is accompanied with the fragmentation of the
   outer ejecta that is possibly related to the seed inhomogeneities generated
   by a vigorous convection in the RSG envelope.
It is remarkable that the early spectra of the peculiar type-IIP SN~1987A
   do not reveal the \Ha/\Hb problem \citep{UC_2013} which is consistent with
   the absence of the vigorous convection in the envelope of the blue supergiant,
   the progenitor of SN~1987A.

The early-time boundary velocity of the ejecta combined with the \Ha HVNA permits
   us to recover the CS wind density that turns out to be typical for the RSG with
   a moderate mass-loss rate.
The important implication of the found CS density is that the CS interaction
   luminosity is significantly lower compared to the initial bolometric
   luminosity of SN~2018gj.
It thus rules out a notable contribution of the CS interaction to the
   luminosity primarily related to the radiative cooling of the exploded RSG
   with the radius of 775\Rsun.

The recovered ejecta mass together with the mass of the collapsing core of 1.6\Msun
   suggests the pre-SN mass of 25\Msun.
The inferred pre-SN mass-loss rate $\dot{M} = 2\times10^{-6}u_{15}$\Msyr
   is consistent with the data for eight 25\Msun RSG showing $\dot{M}$
   in the range of $(0.2-5.6)\times10^{-6}$\Msyr \citep{Beasor_2020}.
The found mass-loss rate implies that the progenitor lost $\sim$1.5\Msun
   at the RSG stage.
With the mass of $\sim$3\Msun lost earlier via a high-velocity wind at
   the hydrogen-burning stage \citep{Beasor_2020} one expects that
   the SN~2018gj initial mass on the main sequence was about 29\Msun.

\begin{figure}
   \includegraphics[width=\columnwidth, clip, trim=8 23 28 28]{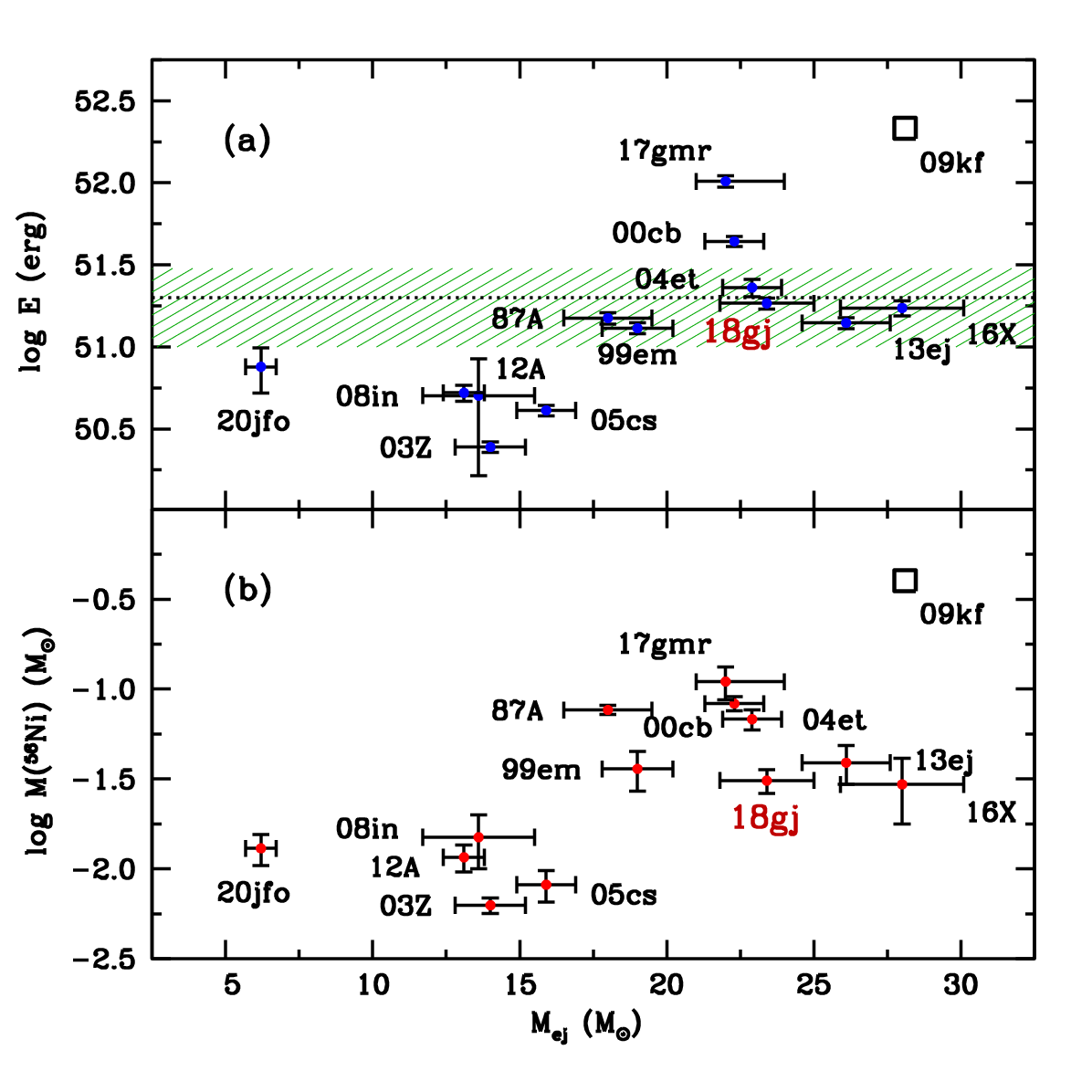}
   \caption{
   Explosion energy (Panel (a)) and $^{56}$Ni mass (Panel (b))
      versus ejecta mass for SN~2018gj and 13 other
      core-collapse SNe studied using the uniform approach \citep{UC_2024}.
   The \emph{dotted line} in Panel (a) is the upper limit of the explosion energy of
      $2\times10^{51}$\,erg for the neutrino-driven mechanism \citep{Janka_2017}
      with the uncertainty of about $\pm 10^{51}$\,erg\protect\footnotemark\
      shown by the shaded \emph{green band}.
   }
   \label{fig:ennims}
\end{figure}
\footnotetext{H.-Th.~Janka, private communication.}  
The number of SNe~IIP explored via a unified hydrodynamic modeling now amounts
   to 14 (Table~\ref{tab:sumtab}).
Their properties are illustrated on the scatter diagrams: the explosion energy vs.
   the ejecta mass and the $^{56}$Ni mass vs. the ejecta mass
   (Fig.~\ref{fig:ennims}).
SN~2018gj does not deviate from the general behavior of the explosion energy and
   the $^{56}$Ni mass versus the ejecta mass.
 
A remarkable feature of SN~2018gj is the high velocity of the $^{56}$Ni ejecta.
We interpret the significant blueshift of the \Ha emission at the nebular epochs
   \citep{Teja_2023} as an effect of the aspherical $^{56}$Ni ejecta, being
   possibly of a bipolar geometry, likewise in SN~2004dj \citep{Chugai_2005}.
Our sample of SNe~IIP (Table~\ref{tab:sumtab}) includes the events with
   an apparent $^{56}$Ni asymmetry (7 cases) and with the seemingly symmetrical
   $^{56}$Ni ejecta (6 events).
We deliberately exclude SN~2009kf from the latter sample, because this object
   has the extremely high explosion energy $\approx$2$\times10^{52}$\,ergs that
   indicates a hypernova explosion mechanism rather than the neutrino-driven
   explosion of SNe~IIP with the energy of $\lesssim$2$\times10^{51}$\,erg
   \citep{Janka_2017}.

It is remarkable that SNe~IIP with the symmetric $^{56}$Ni in our sample have
   predominantly lower velocities of the $^{56}$Ni ejecta compared to
   SNe~IIP with the apparent asymmetry.
Indeed, the mean of the $^{56}$Ni outer velocity for six symmetric SNe~IIP is
   $890\pm340$(s.d.)\kms, whereas for seven asymmetric SNe~IIP the mean is
   $4440\pm2340$\kms.
It should be noted that in the case of the low $^{56}$Ni velocity its asymmetry
   is difficult to detect.
The case of SN~1999em with the low $^{56}$Ni velocity and the apparent bipolar
   \Ha asymmetry \citep{Chugai_2007a} is an exception, possibly due to the
   $^{56}$Ni bipolar axis being colinear with the line of sight.
One can assume, therefore, that the aspherical $^{56}$Ni ejecta is an intrinsic
   feature of all SNe~IIP.

\section{Conclusions}
\label{sec:concl}
We conclude with a summary of major results:
\begin{itemize}
\item%
The hydrodynamic modeling of type-IIP SN~2018gj with the short plateau suggests
   the explosion of a $\approx$25\Msun RSG star with a radius of
   $\approx$775\Rsun that ejects $\approx$23.4\Msun with the energy of
   $\approx$1.8$\times10^{51}$\,erg.
\item%
The analysis of hydrogen lines in the early spectra reveals two major facts:
   (i) a high velocity of the outer ejecta (15\,000\kms) that is inconsistent
   with the presence of a confined massive CS shell;
   (ii) a clumpy structure of the outer ejecta of SN~2018gj and SN~2020jfo.
\item%
The recovered wind density rules out a noticeable contribution of
   the CS interaction to the bolometric luminosity.
\item%
The significant early escape of gamma-quanta suggests a high velocity
   ($\approx$5200\kms) of the $^{56}$Ni ejecta.
\item%
The analysis of the sample of 13 SNe~IIP studied hydrodynamically in a uniform way
   implies that the asymmetry of the $^{56}$Ni ejecta may be an intrinsic feature of
   all SNe~IIP.
\end{itemize}

\backmatter

\bmhead{Acknowledgements}
Not applicable.

\bmhead{Author contribution}
V.U. and N.C. contributed equally to this work.

\bmhead{Funding}
Not applicable.

\bmhead{Data availability}
No datasets were generated or analyzed during the current study.

\bmhead{Materials availability}
Not applicable.

\bmhead{Code availability} 
Not applicable.

\section*{Declarations}

\subsection*{Ethics approval and consent to participate}
Not applicable.

\subsection*{Consent for publication}
Not applicable.

\subsection*{Competing interests}
The authors declare no competing interests.


\end{document}